\documentclass[aps,nofootinbib,preprintnumbers,twocolumn,superscriptaddress,prd,twocolumn]{revtex4-1}

\usepackage{amsmath}
\usepackage{amssymb}
\usepackage{slashed}
\usepackage{epsfig}
\usepackage{bbm,bm}
\usepackage{hyperref}
\usepackage{color}
\usepackage{comment}
\usepackage{amsfonts}
\usepackage{dsfont}
\usepackage{multirow}
\usepackage{tensor}
\usepackage[normalem]{ulem}
\usepackage{xcolor,colortbl}
\usepackage{etex}
\usepackage{braket}
\usepackage{float}
\usepackage{slashed}
\usepackage{xcolor,colortbl}
\usepackage{soul}
\usepackage{nicefrac}
\usepackage{dcolumn} 
\usepackage{tikz}

\usepackage{mathpazo}

\def\bea{\begin{eqnarray}} 
\def\eea{\end{eqnarray}}
\def\be{\begin{equation}} 
\def\ee{\end{equation}} 
\def\ba{\begin{array}}
\def\ea{\end{array}}

\let\oldtitle\title
\renewcommand{\title}[1]{\oldtitle{\color{blue}{#1}}}

\begin{document}

\title{A multicritical Landau-Potts field theory}

\author{A.\ Codello}
\email{a.codello@gmail.com}
\affiliation{Instituto de F\'isica, Faculdad de Ingenier\'ia, Universidad de la Rep\'ublica, 11000 Montevideo, Uruguay}

\author{M.\ Safari}
\email{mahsafa@gmail.com}
\affiliation{Romanian Institute of Science and Technology, 
Str.~Virgil Fulicea  3, 400022 Cluj-Napoca, Rom\^ania}

\author{G.\ P.\ Vacca}
\email{vacca@bo.infn.it}
\affiliation{INFN - Sezione di Bologna, via Irnerio 46, I-40126 Bologna, Italy}

\author{O.\ Zanusso}
\email{omar.zanusso@unipi.it}
\affiliation{Universit\`a di Pisa and INFN - Sezione di Pisa, Largo Bruno Pontecorvo 3, I-56127 Pisa, Italy}

\begin{abstract}
We investigate a perturbatively renormalizable $S_{q}$ invariant model with $N=q-1$ scalar field components
below the upper critical dimension $d_c=\nicefrac{10}{3}$.
Our results hint at the existence of multicritical generalizations of the critical models of spanning random clusters
and percolations in three dimensions.
We also discuss the role of our multicritical model in a conjecture that involves the separation of first and second order
phases in the $(d,q)$ diagram of the Potts model.
\end{abstract}

\maketitle

\section{Introduction}\label{sect:introduction}

Universality classes characterized by a single real order parameter are generally well understood,
both qualitatively and quantitatively, for all dimensions between $d=2$
and the upper critical dimension $d_c$ of the underlying microscopic model.
The understanding is often based on a combination of renormalization group (RG)
and conformal field theoretical (CFT) methods. The general picture regarding universality classes
with several order parameters, which would correspond to multi-field scalar theories, is instead lacking.
Borrowing some inspiring words from Ref.~\cite{Osborn:2017ucf}, we do not yet have a {\it mappa mundi} giving us a bird's eye view
of the whole spectrum of equilibrium critical phenomena,
even if this field of theoretical physics is hardly a {\it terra incognita} due to decades of research on the topic.

We do know, however, that symmetry must play a crucial role when charting the atlas of critical phenomena.
This can be seen in a multitude of ways, but from our perspective it is most interesting
to point out two specific examples related to the $\epsilon$-expansion.
By solving the most general RG fixed point equations in $d=4-\epsilon$ \cite{Codello:2020lta}
and $d=6-\epsilon$ \cite{Codello:2019isr} for a set number
of scalar fields $N$, it can be shown that solutions emerge with a definite symmetry content $G$,
which is, by construction, a subgroup of the maximal symmetry group, $O(N)$ \cite{Rychkov:2018vya}.

Furthermore, RG deformations at a FP almost always arrange as irreducible representations
of the symmetry group $G$, because the action of the group $G$ commutes with the generator of the dilatations,
consequently characterizing the labels of the spectrum of the underlying CFT.
Sometimes, when this does not happen, logarithmic terms can be produced and the more general
framework of logarithmic CFT (log-CFT) must be introduced to accommodate the changes, in
a way that is going to be relevant later on in the paper.
In a natural way, log-CFTs can be obtained as special parametric limits of
standard CFTs \cite{cardy-log-cft}. This is achieved also within RG methods \cite{Zinati:2020xcn,Safari:2020eut}.

In this paper, we concentrate on one of the most recurring and important symmetry groups,
the permutation group $S_q$, which is a subgroup of $O(N)$ for $N=q-1$ \cite{Amit1976}.
The group can be seen as representing the invariance of a regular $q$-symplex, the hyper-tetrahedron,
embedded in $\mathbb{R}^N$, and therefore is relevant for the description of microscopic crystal models
of the same symmetry.
Landau-Potts field theories with $S_q$ symmetry are well known in $d=6-\epsilon$, in which the model
displays a cubic interaction, and in $d=4-\epsilon$, in
which the model has a quartic interaction that enhances the symmetry by a global reflection $S_q \times \mathbb{Z}_2$.
The most important one is certainly the model in $d=6-\epsilon$, because it is known
to belong to the universality class of the lattice Potts model. This implies that the limits $q\to 0$ and
$q\to 1$ reproduce the universality classes of critical spanning clusters (trees and forests)
and of percolations (both bond and site) \cite{Jacobsen:2003qp,Deng:2006ur}, respectively.

In these regards, our point of view is that there can be many more examples of Landau-Potts field theories,
and that their nature can be uncovered by appropriately using the $\epsilon$-expansion,
while changing the upper critical dimension. The first multicritical generalization of the critical Landau-Potts
field theory in $d=6-\epsilon$, which has also genuine $S_q$ symmetry, is a field theory
with quintic interactions and perturbative expansion constructed in $d=\nicefrac{10}{3}-\epsilon$ dimensions.
This happens because the theory with quartic interaction has its symmetry group enhanced by a global parity,
and therefore it is not a natural multicritical generalization.
In this paper, we construct the quintic model, discuss its critical properties, and assess its relevance
as a multicritical partner of the Potts model.
In so doing, the $\epsilon$-expansion becomes a primary tool to search and discover universality classes.
As long as we are not interested in precise quantitative estimates of critical exponents,
the $\epsilon$-expansion, in a way, overcomes the traditional limitations of perturbation theory by changing the critical dimension
at which it is performed.

The paper is organized as follows:
In Sect.~\ref{sect:model}, we introduce the multicritical $S_q$-symmetric Landau-Potts field theory and discuss its renormalization.
In Sect.~\ref{sect:fp-analysis}, we discuss the RG fixed points as functions of $q$.
In Sects.~\ref{sect:percolation} and Sect.~\ref{sect:forest}, we analyze more carefully the limits $q=1$ and $q=0$, respectively.
In Sect.~\ref{sect:conclusions}, we draw our main conclusions by giving a physical
interpretation to the multicritical point.
We include in appendix~\ref{sect:relevant-formulas} all the relevant RG formulas,
including critical exponents and $\gamma$-functions for composite operators.
We want to stress that all the results listed in the appendix,
even though they have been deferred to the end of the paper to avoid overburdening the main discussion,
are a central part of our work.

\section{The multicritical model in $d_c = \nicefrac{10}{3}$}\label{sect:model}

General Landau-Potts field theories can be constructed by placing a scalar field component
on each vertex of a regular $q$-symplex, which realizes the permutation group as the
subset of rotations, $S_q \subset O(N)$, that leave the symplex invariant.
We normalize the vertices $e^\alpha$ of the $q$-symplex as
\begin{equation}\label{eq:vertices}
\begin{split}
 &\sum_{i=1}^N e_i^\alpha e_i^\beta = q\delta^{\alpha\beta}-1 \,, \qquad
 \sum_{\alpha=1}^{q} e_i^\alpha = 0 \,, \\
 &\sum_{\alpha=1}^{q} e_i^\alpha e_j^\alpha = q\delta_{ij} \,,
\end{split}
\end{equation}
The order parameter is defined as $\psi^\alpha = \sum_i \phi_i e_i^\alpha $, where we introduced explicitely the field components $\phi_i$.
The symmetry group acts on $\phi_i$ as a subgroup of $O(N)$, while on $\psi^\alpha$
as the permutation group of the labels $\alpha=1,\cdots,q$.
A manifestly invariant action can be constructed through monomials containing any power of $\psi^\alpha$
and summing over the label $\alpha$ itself.
To express the monomials in terms of the fields $\phi_i$, it is convenient to introduce the tensors of arbitrary rank
\begin{equation}\label{eq:q-tensors}
\begin{split}
 q^{(n)}_{ij\cdots k} = \frac{1}{q}\sum_{\alpha=1}^q e_i^\alpha e_j^\alpha \cdots e_k^\alpha\,.
\end{split}
\end{equation}
We define the invariant action 
\begin{equation}\label{eq:action}
\begin{split}
 S[\phi] & = \int {\rm d}^d x \Bigl\{
 \frac{1}{2}(\partial\phi)^2
 + V(\phi) \Bigr\}\,,\\
V(\phi) & = \frac{1}{5!} \sum_{ijklm} \Bigl(
 {u} \, \delta_{(ij} q^{(3)}_{klm)} 
 +{v} \, q^{(5)}_{ijklm}\Bigr)\phi_{i}\phi_{j}\phi_{k}\phi_{l}\phi_{m} \,,
\end{split}
\end{equation}
where round parenthesis imply a full symmetrization of the enclosed indices.
The invariance of $S[\phi]$ on $S_q$ is easily proven by inserting the definition \eqref{eq:q-tensors}
and noticing that it is actually a scalar function of powers and derivatives of $\psi^\alpha$.
The potential $V(\phi)$ is the most general quintic singlet function of $\phi_i$.

In the potential of \eqref{eq:action}, we introduce the two couplings ${u}$ and ${v}$ of the model.
In $d=\nicefrac{10}{3}$ dimensions, the action \eqref{eq:action} is perturbatively renormalizable in powers of the couplings.
The leading order RG flow for the potential $V(\phi)$ is known in general, meaning with no restriction of an underlying symmetry,
in a procedure that has been given the name of
functional perturbative RG \cite{ODwyer:2007brp,Osborn:2017ucf,Codello:2017hhh},
which is also strongly tied to a perturbative CFT approach~\cite{Rychkov:2015naa,Codello:2017qek,Codello:2018nbe,Codello:2019vtg}.
It can be obtained from the single component flow, shown in \cite{Codello:2017epp},
and introducing ``flavor'' indices appropriately, as done in \cite{Zinati:2019gct,Zinati:2020xcn}.
The result can be expressed diagrammatically
\begin{equation} \label{eq:flow-potential}
\begin{split}
        \beta_V  &= 
        \frac{3}{4} ~ \, \,
        \begin{tikzpicture}[baseline=-.1cm]
        \draw (0,0) circle (.5cm);
        \draw (-.5,0) to [out=30,in=150] (.5,0);
        \draw (-.5,0) to[out=-30,in=-150] (.5,0);
        \filldraw [gray!50] (0,.5) circle (2pt);
        \draw (0,.5) circle (2pt);
        \filldraw [gray!50] (.5,0) circle (2pt);
        \draw (.5,0) circle (2pt);
        \filldraw [gray!50] (-.5,0) circle (2pt);
        \draw (-.5,0) circle (2pt);
        \end{tikzpicture}
        ~ - \frac{27}{8} ~
        \begin{tikzpicture}[baseline=-.1cm]
        \draw (1.5,0) circle (.5cm);
        \draw (1.5,.5) to[out=-90,in=30] (1.067,-.25);
        \draw (1.5,.5) to[out=-90,in=150] (1.933,-.25);
        \filldraw [gray!50] (1.5,.5) circle (2pt);
        \draw(1.5,.5) circle (2pt);
        \filldraw [gray!50] (1.933,-.25) circle (2pt);
        \draw (1.933,-.25) circle (2pt);
        \filldraw [gray!50] (1.067,-.25) circle (2pt);
        \draw (1.067,-.25) circle (2pt);
        \end{tikzpicture} ~ ,
        \\
       \gamma_{ij} & = 
        \frac{3}{80} ~
        \begin{tikzpicture}[baseline=-.1cm]
        \draw (.5,0) -- (1.,0);
        \draw (-.5,0) -- (-1.,0);
        \draw (0,0) circle (.5cm);
        \draw (-.5,0) to [out=30,in=150] (.5,0);
        \draw (-.5,0) to[out=-30,in=-150] (.5,0);
        \filldraw [gray!50] (.5,0) circle (2pt);
        \draw (.5,0) circle (2pt);
        \filldraw [gray!50] (-.5,0) circle (2pt);
        \draw (-.5,0) circle (2pt);
        \draw[] (1.,0) node[above] { ${\scriptstyle j}$};
        \draw[] (-1.,0) node[above] { ${\scriptstyle i}$};
        \end{tikzpicture} ~ ,
     \end{split}
\end{equation}
in which vertices stand for derivatives of the potential w.r.t.\ the fields and lines correspond to summations
over the flavor indices. The $\gamma$-matrix is the anomalous dimension matrix, which, upon diagonalization, yields
half of the anomalous dimension $\eta$. For this symmetry group, the matrix is already diagonal at critical points
and there is only one anomalous dimension $\eta$,
therefore $\gamma_{ij} = \delta_{ij}\frac{\eta}{2}$.
The diagrams of \eqref{eq:flow-potential} are not Feynman diagrams, but the loop count that they display agrees with the underlying
diagrammatic computations, that is obtained by renormalizing three loop contributions.

\begin{widetext}

Inserting the potential \eqref{eq:action} in \eqref{eq:flow-potential} and iteratively
simplifying long strings of products of $e_i^\alpha$ with \eqref{eq:vertices}
gives the anomalous dimension
\begin{equation}\label{eq:eta}
\begin{split}
  \eta &=
  \frac{1}{300} {u}^2 (q-2) (q+5)
  +\frac{1}{15} {u} {v} (q-2) (q-1)
  +\frac{1}{30} {v}^2 (q-2) \left(q^2-2 q+2\right) \,,
\end{split}
\end{equation}
and the beta functions
\begin{equation}\label{eq:beta-functions}
\begin{split}
 \beta_{{u}} &=
  -\frac{3 \epsilon }{2}{u} 
  -\frac{3}{200} {u}^3 \left(17 q^2+799 q-4326\right)
  +\frac{1}{5} {u}^2 {v} \left(-97 q^2+726 q-1061\right)
  \\&
  +\frac{1}{4} {u} {v}^2 \left(-25 q^3+256 q^2-714 q+1156\right)
  +\frac{10}{3} {v}^3 \left(2 q^2+21 q-48\right)\,,
  \\
 \beta_{{v}} &=
  -\frac{3  \epsilon }{2}{v}
  -\frac{3}{25} {u}^3 (5 q+139)
  +\frac{1}{40} {u}^2 {v} \left(-25 q^2-3483 q+7546\right)
  \\&
  +\frac{1}{2} {u} {v}^2 \left(-217 q^2+807 q-890\right)
  +\frac{1}{12} {v}^3 \left(-459 q^3+2296 q^2-4674 q+3756\right)\,.
\end{split}
\end{equation}
We have included the scaling terms proportional to $\epsilon$ by going to
$d=\nicefrac{10}{3}-\epsilon$ dimensions.
The $\epsilon$-expansion can be obtained by solving $\beta_{{u}}=\beta_{{v}}=0$ perturbatively in powers of $\epsilon$
and inserting the solution in critical exponents such as $\eta$.

\end{widetext}

It is clear that $q$ can be analytically continued in the above formulas, so it does not necessarily need to be a positive natural number
bigger than one. This is particularly useful considering that lattice models with $S_q$ symmetry can
also be continued to arbitrary values of $q$ thanks to the Fortuin-Kasteleyn representation \cite{Fortuin:1971dw}.
This is particularly relevant for the limits $q\to 1$ and $q\to 0$, which are central in the theory of random cluster models.

Several more RG related quantities can be computed from \eqref{eq:flow-potential} in the functional framework, as discussed in some detail in \cite{Safari:2020eut}.
All the RG quantities that we have computed can be found in appendix~\ref{sect:relevant-formulas}.
Arbitrary composite operators ${\cal O}(\phi)$ can be introduced by coupling them to an appropriate source ${\cal J}_{\cal O}$ in the path-integral
through the replacement $ S[\phi] \to S[\phi] + {\cal J}_{\cal O}\cdot {\cal O}(\phi)$, in which ${\cal J}_{\cal O}$ renormalizes multiplicatively
when computing $\langle {\cal O}(\phi)\rangle$.
In general, ${\cal J}_{\cal O}$ mixes with other sources, unless ${\cal O}(\phi)$ is already a scaling operator.
The operators ${\cal O}(\phi)$ that we consider are relevant (in the RG sense) and built from powers of $\phi_i$ with no derivatives.
Specifically, in this manner we include in this work the complete spectrum of symmetric operators that are quadratic or cubic in $\phi_i$,
and the scalar operators that are quartic. For obvious reasons, the operators must carry a representation label for $S_q$,
which comes from the tensor product of standard (vector) representations of $S_q$.

For arbitrary values of $q$, the action of $S_q$ and of the generator of dilatations commute,
resulting in some scaling operators which carry an irreducible representation (irrep) label of $S_q$ \cite{Vasseur:2013baa,Couvreur:2017inl}.\footnote{
The space of cubic irreps operators is smaller than
the one generated by $\phi_i\phi_j\phi_k$, as opposed to the quadratic case.
In $d=6-\epsilon$, it was observed, using the results of \cite{Codello:2019isr},
that operators that break $S_q$ and cannot be arranged in terms of irreps have critical exponents with $\epsilon$-expansion starting with ${\mathsf{O}}(\epsilon^2)$, even after a global $O(N)$ rotation has been factored out \cite{Safari:2020eut}.
}
A list of quadratic and cubic operators that we consider here is already given in \cite{Safari:2020eut}, so we simply summarize it briefly
to clarify the notation, but omit most of the long explicit expressions for brevity.
With two copies of $\phi^i$, we can construct a singlet $S^{(2)}=\phi^2$, a vector $V^{(2)}_i \sim \sum_{jk} q_{ijk}\phi_j\phi_k$,
and a symmetric tensor $T^{(2)}_{ij}$. Similarly, with three copies of $\phi^i$,
we can construct a singlet $S^{(2)}=\sum_{ijk} q_{ijk}\phi_i\phi_j\phi_k$,
two vectors $V^{(3)}_i $ and $V^{(3)\prime}_i $, a symmetric tensor $T^{(3)}_{ij}$, and a symmetric $3$-tensor $Z^{(3)}_{ij}$.
We also include the two scalar singlets $S^{(4)}$ and $S^{(4)\prime}$ coming from four copies of $\phi^i$.
We refer to appendix~\ref{sect:relevant-formulas} for explicit forms of the mixing cubic vectors and quartic scalars,
while all other operators are given explicitly in \cite{Safari:2020eut}.

For each scaling operator ${\cal O}$, the renormalization process introduces a $\gamma$-function
and consequently a critical exponent $\theta_{\cal O}$,
which can be related to the CFT operator scaling dimension as $\Delta_{\cal O}=d-\theta_{\cal O}$
(if ${\cal O}$ is primary, if not this formula is slightly modified to accommodate $\Delta_{V'}=2+\Delta_{\phi}$).

The physical meaning of the critical exponents has to do with observable quantities at criticality,
in agreement with their ``quantum'' numbers.
For example, the leading critical behavior of the energy is governed by the scaling of the first singlet $\theta_{S^{(2)}}$,
and all subleading corrections are given by $\theta_{S^{(n)}}$ for $n\geq 3$.
Similarly, the leading critical behavior of the magnetization is governed by  $\theta_{V^{(1)}}=(d+2-\eta)/2$
and subleading corrections are $\theta_{V^{(n)}}$ for $n\geq 2$.

For increasing rank of the operators, also the number of available operators increases, and at any order there is
a correction to the energy and the magnetization.
Further symmetric $n$-tensors describe the ``propagation'' of $n$-clusters at criticality \cite{Vasseur:2013baa}.
For some special limits of $q$, there can be a degeneracy in the spectrum in which two operators transforming with a
different irrep have the same scaling dimension in the limit. This results in
a non-diagonalizable CFT Jordan cell \cite{cardy-log-cft,Hogervorst:2016itc}
and, consequently, in a universal logarithmic correction to an operator that mixes the original two \cite{Vasseur:2013baa,Safari:2020eut}.
The only relevant case for this paper happens when $q\to1$, in which $\Delta_{S^{(2)}}=\Delta_{T^{(2)}}$.
In this case, relevant for percolations in the standard universality class, the space of tensors actually has negative dimension;
if seen as an analytic continuation, the tensors disappear from the spectrum by ``colliding'' with the singlets.
Notably, an observable displaying this logarithmic behavior can be explicitly constructed \cite{vjs}
and the coefficient $\alpha$ of the logarithm
can be computed from $\Delta_{T^{(2)}}-\Delta_{S^{(2)}} \sim \alpha (q-1) + {\mathsf{O}}(q-1)^2$.
An explicit form of the logarithmic correlator is given at the end of appendix~\ref{sect:relevant-formulas}
and is discussed in much more detail in \cite{vjs}.

\section{General fixed points analysis}\label{sect:fp-analysis}

We solve $\beta_{{u}}=\beta_{{v}}=0$ for the system \eqref{eq:beta-functions} and an arbitrary value $q$.
The solution exists analytically, but it is best displayed numerically due to its complexity.
We study all the possible real non-Gaussian solutions as functions of $q$, for $q\geq 0$ and modulo reflections
$({u},{v}) \leftrightarrow -({u},{v})$,
so we can include all the interesting limits $q=0,1,\cdots$ and so on.
The system is best analyzed by first rescaling $\epsilon$ away,
through the definitions ${u}= a \sqrt{\epsilon}$ and ${v}= b \sqrt{\epsilon}$.
The fixed point equations become $f_q(a,b)=\epsilon^{-\nicefrac{3}{2}}\beta_{{u}}=0$
and $g_q(a,b)=\epsilon^{-\nicefrac{3}{2}}\beta_{{v}}=0$.
Using the rescaled couplings $a$ and $b$, fixed points are thus solutions of two cubic equations.

A particularly elegant way to find all possible real solutions for a given value of $q$
involves the use of a "triangular" Gr\"obner basis for the polynomials in the rescaled couplings $(a,b)$.
Using this basis, an equation for the coupling $b$ can be found as its first element,
and it takes the form $bP_q(b)=0$, with $P_q(b)$ being an even polynomial of eighth order (having factored out the Gaussian solution).
The other equation is of the form $a=b R_q(b)$, with $R_q(b)$ being an even polynomial of sixth order.
We do not give the explicit form of $P_q(b)$ here, because it is rather long,
but we provide it for the special cases $q=1$ and $q=0$
in Sects.~\ref{sect:percolation} and~\ref{sect:forest}, respectively.
Notice that the Gaussian solution is the only solution with zero couplings.
The number of real zeroes of $P_q(b)$ changes according with $q$
and, as it should be evident, it is in one-to-one correspondence with
the nontrivial solutions of the full system.

\begin{center}
\begin{figure*}[t!]
\includegraphics[width=0.8\textwidth]{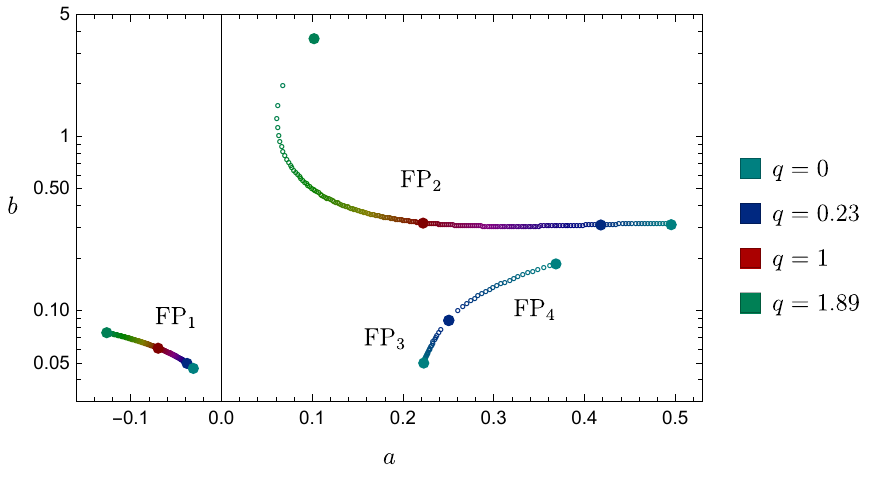}
\caption{
Fixed points for varying $q$. The fixed points ${\rm FP}_1$ and ${\rm FP}_2$, respectively located lower-left and upper-right,
move roughly from right to left for increasing $q$. We also have that ${\rm FP}_2$ goes to infinity for $q=q_2\approx 1.89$.
The fixed points ${\rm FP}_3$ and ${\rm FP}_4$ start by moving towards each other, and then merge at $q=q_1\approx 0.23$.
The vertical axis is displayed on a logarithmic scale.
\label{fig:fpmap}
}
\end{figure*}
\end{center}

There are three special values of $q$, which we can give numerically
\begin{equation}
\begin{split}
  q_1 = 0.2304\,, \qquad q_2=1.8940\,, \qquad q_3=3.8778\,.
\end{split}
\end{equation}
For $0\leq q < q_1$ there are four distinct fixed points (modulo reflections), labelled ${\rm FP}_i$ for $i=1,\cdots,4$.
Their location in the $({u},{v})$ roughly mimic the RG diagram of the special case $q=0$,
which is discussed in more detail in Sect.~\ref{sect:forest}.
Crossing $q=q_1=0.2304$ we see that ${\rm FP}_3$ and ${\rm FP}_4$ collide,
consequently for $q_1 < q < q_2$ there are two distinct fixed points.
In this second case, the RG diagram in the place $({u},{v})$ resembles the one of the special case $q=1$,
which is discussed in Sect.~\ref{sect:percolation}.
Crossing $q=q_2=1.8940$ we see that ${\rm FP}_2$ goes to infinity,
so for $q_2 < q < q_3$ there is only one fixed point.
Finally, crossing $q=q_3=3.8778$ the last fixed point also goes to infinity
and there are no solutions.
The $q$-dependent behavior of the fixed point solutions is shown in Fig.~\ref{fig:fpmap}.

There are four natural values of $q$ contained in the interval $0\leq q < q_3$,
which could, in principle, lead to interesting critical points.
The cases $q=0$ and $q=1$ are in fact interesting, and we study them in the respective sections below.
The case $q=2$ is probably less interesting: the limit of $2$-states
reduces the symmetry to the one of an Ising ferromagnet, $S_2\simeq \mathbb{Z}_2$,
which cannot be realized through an odd-potential (at most there can be a conjugation parity as in \cite{Codello:2017epp}).
This can be seen in two ways: on the one hand the critical exponents are Gaussian for $q\to 2$ as evident from \eqref{eq:eta},
on the other hand the potential itself is trivial, $V(\phi)=0$ for $q=2$,
when expressing it with the basis $e^{\alpha}=\pm 1$ for $\alpha=1,2$ respectively.
Finally, the case $q=3$ is also trivial, but in a slightly less straightforward way.
For $q=3$, the two quintic invariants of \eqref{eq:action} coincide if explicitly computed,
resulting in a potential $V(\phi)\propto (2{u}+5{v})(\phi^2_1+\phi_2^2) \phi_2(\phi_2^2-3\phi_1^2) $,
but at the fixed point $2{u}+5{v}=0$, even though the two couplings have
nonzero values. Consequently, $V(\phi)=0$ also for $q=3$ and there is only the Gaussian solution.

Nevertheless,
we notice that ${\rm FP}_1$ still exists in the interval containing $q=2$ and $q=3$, upon analytic continuation.
In the first case, it could imply the presence of an interesting $\mathbb{Z}_2$ model,
realized as one with $S_q$ symmetry with $q=2+\varepsilon$ for $\varepsilon\to 0$,
however, it would have $\eta=0$ from \eqref{eq:eta}.
Therefore, the above analysis does not exclude the possibility that the cases $q=2$ and, to some extent, $q=3$ are entirely trivial.
For example, they could, in fact, still produce logarithmic corrections in the way discussed in Sect.~\ref{sect:introduction},
and would represent Gaussian theories with some logarithmic correlators \emph{if opportune observables
are found}, similarly to what has happened for the case $q=1$ in \cite{vjs} (which, however, is nonGaussian).
Since we do not know of interesting observables of this type yet, we take the analysis of this
section as an indication that the interesting limits of this model that deserve a more careful analysis
are $q=0$ and $q=1$, that incidentally are related to the two most important random cluster models
(at least to our eyes).

\section{The limit $q\to 1$: percolations}\label{sect:percolation}

The limit $q\to 1$ for microscopic random cluster models is known to be related to the universality class of percolations.
In fact, the Landau-Potts field theory with cubic interaction in $d=6-\epsilon$ dimension is known to belong
to the same univarsality class as bond and site percolations.
When applying the same limit to our multicritical model, we can argue that our findings suggest
the existence of a multicritical generalization of the standard percolation universality class.
One way to think at the generalization is to recall how the standard Ising model,
that has upper critical dimension $d_c=4$, is generalized to the tricritical Ising model,
that has $d_c=3$, by including a new $\mathbb{Z}_2$ relevant parameter.
Since our action \eqref{eq:action} has $d_c=\nicefrac{10}{3}$, we expect that the model is nonGaussian
in $d=3$ dimensions given that the required $\epsilon=\frac{1}{3}\lesssim 1$ for the continuation is relatively small,
unless the perturbative expansion fails rather miserably, that we have no reason to believe.
Real-world multicritical generalization of percolations appear, for example,
in the critical behavior of correlated percolation \cite{Delfino:2009rq,Delfino:2010af}.

\begin{figure}
\begin{center}
\includegraphics[width=7.5cm]{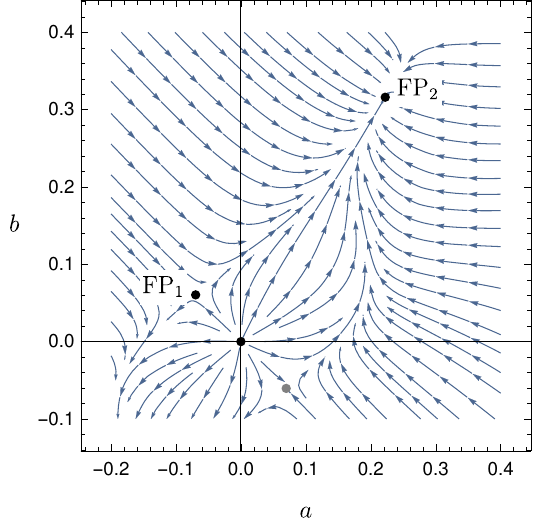}
 \caption{RG flow in the limit $q\to 1$. The arrows point towards the infrared and marked dots are the fixed points (and their mirror images).
  Notice that there are trajectories connecting the Gaussian fixed point and ${\rm FP}_1$ to ${\rm FP}_2$ in the infrared.
 \label{fig:plot-q=1}
 }
\end{center}
\end{figure}

The first polynomial of the Gr\"obner basis for $q\to 1$ is
\begin{equation}
\begin{split}
 P_{q=1}(b) &= 87480 (
  89038468249947 b^8-8972429711878 b^6
 \\
 &+12888105748 b^4-536171880 b^2+2211840)\,.
\end{split}
\end{equation}
As already hinted at in the previous section, the above polynomial has two real zeroes,
so we actually have two nontrivial fixed points, ${\rm FP}_1$ and ${\rm FP}_2$ (apart from reflections).
We plot them in Fig.~\ref{fig:plot-q=1}. One can clearly see that ${\rm FP}_2$ is more IR relevant than ${\rm FP}_1$,
so it is a more realistic candidate for the multicritical universality class, but for completeness we report
the results for both.
%
%

For ${\rm FP}_1$ we find the anomalous dimension and the critical exponent of the correlation length
\begin{equation}
\begin{split}
  \eta= -0.000219126 \epsilon\,, \qquad
  \nu^{-1} = 2 + 0.00460164 \epsilon\,.
\end{split}
\end{equation}
Geometric properties of the critical clusters are characterized by the fractal and resistivity dimensions
\begin{equation}
\begin{split}
  d_f = 2 + 0.0121628 \epsilon \,, \qquad
  d_r = \nu^{-1} \,,  
\end{split}
\end{equation}
with the latter being determined by a scaling relation.
Crossover properties and logarithmic properties are governed by
\begin{equation}
\begin{split}
  &\Phi=1\,,\qquad \overline{\Phi} = 1 + 0.00378056 \epsilon\,,\\
  &\alpha_E = -0.0069779\epsilon\,,
\end{split}
\end{equation}
and the formulas to obtain the above quantities are defined in appendix~\ref{sect:relevant-formulas}.

For ${\rm FP}_2$ we find the critical exponents
\begin{equation}
\begin{split}
  \eta= -0.00431785 \epsilon\,, \qquad
  \nu^{-1} = 2 + 0.0906748\epsilon\,.
\end{split}
\end{equation}
The fractal dimensions
\begin{equation}
\begin{split}
  d_f = 2 + 0.11886 \epsilon \,, \qquad
  d_r = \nu^{-1} \,,  
\end{split}
\end{equation}
the crossover exponents and logaritmic coefficient
\begin{equation}
\begin{split}
  &\Phi=1\,,\qquad \overline{\Phi} = 1 + 0.0140927  \epsilon\,, \\
  &\alpha_E =-0.0222583 \epsilon\,.
\end{split}
\end{equation}

All critical exponents and properties of both fixed points ${\rm FP}_1$ and ${\rm FP}_2$ in the limit $q\to 1$
are summarized in Tab.~\ref{tab:percolation}, which can be found in App.~\ref{sect:relevant-formulas}.
The fractal dimensions of the critical cluster are probably the most direct geometric consequence of the critical regime.
A naive extrapolation to $d=3$, corresponding to $\epsilon=\frac{1}{3}$, gives $d_f= 2.00405$ and $d_r=2.00153$ for ${\rm FP}_1$,
and $d_f= 2.03962$ and $d_r=2.03022$ for ${\rm FP}_2$. We are clearly talking about very mild departures from
the mean field values, even smaller if $d_f$ and $d_r$ are compared relatively to each other.
The most interesting fixed point ${\rm FP}_2$, however, has also the most significant correction since it affects
the second digit. We hope that this could be a useful signature for finding this critical point in a microscopic model
that generalizes the one of bond percolations with additional tunable parameters.

\section{The limit $q\to 0$: spanning forests}\label{sect:forest}

The limit $q\to 0$ of random cluster models is known to be related to models of spanning clusters, such as trees and forests.
As a consequence, and on the basis of the analogy of the previous case and of the Potts model,
the Landau-Potts field theory at criticality is believed to belong to the same universality class.
The Gr\"obner polynomial in this limit becomes
\begin{equation}
\begin{split}
 P_{q=0}(b) &= 41472 ( 736 b^2-25) (342501160032 b^6
 \\&-34407942288 b^4+153295414 b^2-173889)\,.
\end{split}
\end{equation}
Clearly, $P_{q=0}(b)$ is more factorized than its counterpart of the previous section.
We eliminate reflections of the fixed points through the requirement $b>0$
($b\to -b$ and $a\to -a$ corresponding to completely equivalent solutions).
There are four real zeroes of $P_{q=0}(b)$, one corresponding to the first factor, and the other three corresponding to
the second one.
The first factor is solved by $b=5/4\sqrt{46}$, that corresponds to ${\rm FP}_4$;
${\rm FP}_i$ with $i=1,2,3$ come as solutions to the second.

\begin{figure}
 \begin{center}
\includegraphics[width=7.5cm]{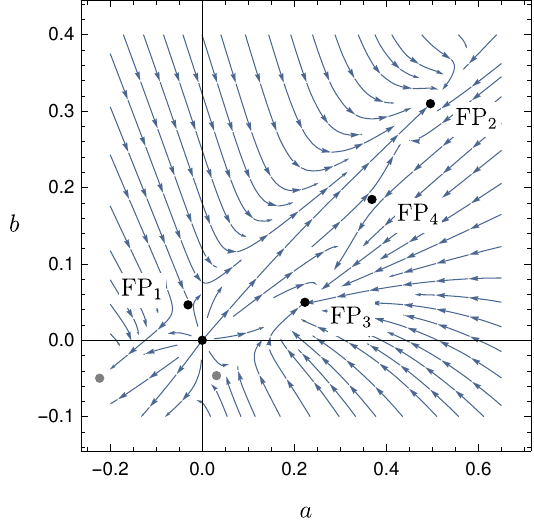}
 \caption{RG flow in the limit $q\to 0$. Arrows point towards the infrared and marked dots are the fixed points (and their mirror images).
 The most infrared stable fixed points are ${\rm FP}_2$ and ${\rm FP}_3$.
  \label{fig:plot-q=0}
  }
 \end{center}
\end{figure}

We plot the solutions in Fig.~\ref{fig:plot-q=0}. One can clearly see that ${\rm FP}_2$ is more IR relevant than ${\rm FP}_1$,
so it is a more realistic candidate for the multicritical universality class, but for completeness we report the results for both.
Before proceeding, we also notice that ${\rm FP}_3$ is also an IR fixed point;
as much as ${\rm FP}_2$, it is completely IR attractive, and both have trajectories
connecting them from ${\rm FP}_2$, ${\rm FP}_4$ and the Gaussian fixed point.

The factorization property comes because there actually is a more convenient coupling to work with, namely the difference
\begin{equation}
\begin{split}
 \chi={u}-2{v}\,.
\end{split}
\end{equation}
The new coupling has an independent beta function (in the sense that it only depends on $\chi$ itself)\footnote{%
This seems to be a property exclusive to the case $q=0$, which does not have an equivalent for the other values that we studied
(at least using a linear combination of the couplings).
It is entirely possible that the operator corresponding to this parametrization has a special physical meaning.
It could also be that the number of couplings should be reduced to one in the limit $q\to 0$, and
this could explain why one fixed points has almost Gaussian critical properties (see later in the section),
since a redundant coupling often results in a duplicate of the Gaussian fixed point.
However, we have not been able to find any special property,
therefore this issue could deserve a deeper investigation using the methods of \cite{Vasseur:2013baa,Couvreur:2017inl}.
}
\begin{eqnarray}
 \beta_{\chi} = -\frac{3}{2}\epsilon \chi + \frac{393}{4}\chi^3\,.
\end{eqnarray}
The only possible solutions of $\beta_{\chi}=0$ are the Gaussian one and $ \chi^* = \pm \sqrt{\nicefrac{2\epsilon}{131}} $.
In the full system of solutions, the Gaussian fixed point and ${\rm FP}_4$ share $\chi=0$,
while ${\rm FP}_i$ for $i=1,2,3$ share $\chi=\chi^*$.

The importance of the new coupling is evident also from the explicit form of the standard critical exponents
\begin{eqnarray}
 \eta = -\frac{\chi^2}{30}\,, \qquad \nu^{-1}=2+\frac{7}{10} \chi^2\,,
\end{eqnarray}
that are thus independent of any orthogonal coupling. This implies that the critical exponents of ${\rm FP}_4$ are trivial,
while ${\rm FP}_i$ for $i=1,2,3$ share the same $\eta$ and $\nu$.
The same happens for the critical exponent of the quadratic vector operator is $ \theta_{V^{(2)}}=2+\frac{6}{5} \chi^2$,
implying that also $d_f$ and $\overline{\Phi}$ are shared by the various fixed points.
However, not all critical exponents depend on $\chi$, for example the quadratic tensor is
$\theta_{T^{(2)}}=2+\frac{ \epsilon }{25} \left(13 {u}^2-20  {u} {v}+20 {v}^2\right)$
and does not only depend on $\chi$.
A way to interpret this structure is that the two groups of fixed points, ${\rm FP}_4$ and the Gaussian on the one hand,
and ${\rm FP}_i$ for $i=1,2,3$ on the other hand, share several critical properties, but not all of them.
For example, the spectrum of ${\rm FP}_4$ looks almost Gaussian, but deviates from Gaussianity when
higher point correlators are considered.

We still think that ${\rm FP}_2$ is the most interesting fixed point, being completely IR attractive.
Considering that ${\rm FP}_1$ and ${\rm FP}_2$ interpolate continuously with the fixed points discussed in the previous section,
we report some of their critical properties. As already noticed, they share the critical exponents
\begin{equation}
\begin{split}
  \eta=  -0.000508906 \epsilon\,, \qquad
  \nu^{-1} = 2 + 0.010687 \epsilon\,,
\end{split}
\end{equation}
the fractal dimension $d_f =2+0.0183206 \epsilon$, and the crossover exponent $\overline{\Phi} = 1 +0.00381679  \epsilon$.
They differ however in the determination of the exponent
\begin{equation}
\begin{split}
  &\Phi = 1 -0.0036637 \epsilon \,, \qquad {\rm for \,\,\, FP}_1 \,,\\
  &\Phi = 1 +0.0354856 \epsilon \,, \qquad {\rm for \,\,\, FP}_2 \,,
\end{split}
\end{equation}
and the resistivity dimension $d_r$
\begin{equation}
\begin{split}
  &d_r = 2 + 0.0033595 \epsilon \,, \qquad {\rm for \,\,\, FP}_1 \,,\\
  &d_r = 2 + 0.0816582 \epsilon \,, \qquad {\rm for \,\,\, FP}_2 \,.
\end{split}
\end{equation}
As for the previous section, ${\rm FP}_2$ gives the most sizable corrections and is more likely to be seen
in a microscopic model.

All critical exponents and properties of the fixed points ${\rm FP}_1$ and ${\rm FP}_2$ in the limit $q\to 0$,
as well as those of the fixed points ${\rm FP}_3$ and ${\rm FP}_4$ that have not been discussed in this section,
are summarized in Tab.~\ref{tab:forest}, which can be found in App.~\ref{sect:relevant-formulas}.

\section{Physical interpretation and discussion}\label{sect:conclusions}

In this paper, we have discussed a multicritical generalization of the Landau-Potts field theory with quintic
interaction that admits a perturbatively renormalizable $\epsilon$-expansion below the upper
critical dimension $d_c=\nicefrac{10}{3}$ and is therefore non trivial in three dimensions.
This model has genuine $S_q$ symmetry like the Landau-Potts theory with cubic interaction,
differently than the quartic hypertetrahedral model, that could also be interpreted as a generalization,
but has symmetry enhanced by a global factor $\mathbb{Z}_2$.

Using the analytic continuation in the number of states $q$ and explicitly evaluating the potential for some natural values
$q\geq 2$, we have observed that the only natural values for which the model has
nontrivial fixed points are $q=0$ and $q=1$. This is an interesting observation, because these two limits,
if applied to a microscopic model in the Fortuin-Kasteleyn representation, lead to
models of spanning random clusters and of percolations for $q=0$ and $q=1$, respectively.
This fact strongly suggests that it should be possible, by opportunely introducing at least one relevant deformation,
to construct a multicritical point in the phase-diagram of the above random cluster models.
There are already multicritical points in the phase diagram of percolations, so this model might
be relevant to discuss them.\footnote{More precisely, there is a multicritical point in the phase diagram of the Ising model,
in which spin clusters become percolating \cite{Delfino:2009rq,Delfino:2010af}.}

\begin{figure}
\includegraphics[width=0.45\textwidth]{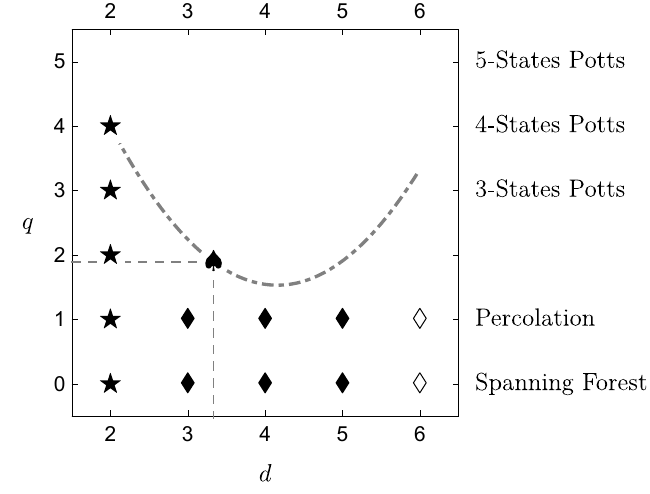}
\caption{
Conjectured depiction of the separation between phase transition orders for the $q$-states Potts universality class as a function of $(d,q)$.
Above the line, whose exact position is unknown, the transition is of first order, while below it is of second order.
Stars ($\bigstar$) correspond to models for which the exact solution is known through CFT methods in $d=2$.
Filled diamonds ($\blacklozenge$) indicate that the model exists and is non-trivial (not mean-field),
while empty diamonds ($\lozenge$) indicate the dimensions for which there is the onset of mean-field critical exponents
(logarithmic corrections to scaling). The special point determined in Sect.~\ref{sect:conclusions} is marked by a spade symbol ($\spadesuit$).
Numerical simulations suggest that the point $(3,3)$ is first order, in agreement with the depicted separatrix.
We intentionally leave out the Ising universality class,
because it is governed by a quartic interaction with $d_c=4$.
\label{fig:diagram}
}
\end{figure}

Here, however, we want to discuss an additional possibility to explain the physical meaning of the multicritical point.
First, we recall that an interesting diagram to study is the ``existence'' diagram for nontrivial critical points
of the lattice Potts model as a function of the dimension $d$ and the number of states $q$.
The fine details of this diagram are not known \cite{Gorbenko:2018ncu,Gorbenko:2018dtm}, but
a rough idea of this diagram can be obtained combining information from CFT in $d=2$,
numerical simulations in $d=3$, and perturbation theory of the Landau-Potts theory with cubic interaction in $d=6-\epsilon$.
This is shown in Fig.~\ref{fig:diagram},
from which we remove the case $q=2$ corresponding to the Ising model because the critical interaction is established to be quartic
and would represent a special case in the diagram \cite{Zia:1975ha}.
Combining all information together, and including a separatrix that visually aids the separation,
the diagram is divided in two parts, roughly corresponding to the top and bottom parts. The bottom part
includes the values of $(d,q)$ with a nontrivial critical point for which we expect
a phase transition of the second order, while the top part includes those without and for which we expect
a first order behavior.

The exact position of the separatrix is not well known. Using CFT information in $d=2$, we know that it must interpolate
with the $4$-states Potts model, thus crossing the point $(2,4)$ in the diagram
\cite{Buffenoir:1992mz,Delfino:2000xt,Gorbenko:2018ncu,Gorbenko:2018dtm}.
Using RG in $d=6-\epsilon$, we know that,
for $q<\frac{10}{3}$ \cite{Amit1976}, there is a real fixed point (modulo reflection, as usual), that can be used to discuss the cases $q=0$ and $q=1$
just like we have done in the previous sections. Instead, in $d=6-\epsilon$ for $q>\frac{10}{3}$ there is only a purely imaginary fixed point,
that would correspond to a complex CFT. The transition between first and second order thus happens at the point $(6,\nicefrac{10}{3})$.
Furthermore, there are numerical simulations of the $3$-states Potts model in $d=3$, which suggest that the model has a weak first order
transition, implying that it is, probably, just above the separatrix.
The separatrix must therefore go through the points $(2,4)$ and $(6,\nicefrac{10}{3})$, but must also pass below and close to $(3,3)$,
although the precise parametrization of the curve is unclear.

\begin{figure}
\includegraphics[width=0.36\textwidth]{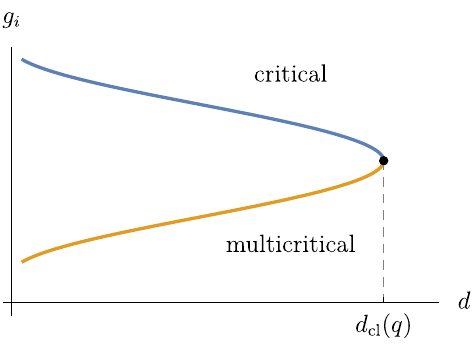}
\caption{
 Illustration of the mechanism for which two fixed points collide. As a function of some couplings $g_i$, a critical and a multicritical
 fixed points exist independently for $d< d_{cl}$, but merge at $d=d_{cl}$ and become a complex conjugate pair above.
 The dimension $d_{cl}$ plays the role of an effective upper critical dimension, to some extent.
 For the universality class of the $q$-states Potts model, we expect the dimension at which they collide to be a function of $q$, $d_{cl}=d_{cl}(q)$,
 according to Fig.~\ref{fig:diagram}. The same collision mechanism is verified for varying $q$ at fixed $d=2$ \cite{Gorbenko:2018dtm}.
\label{fig:merging}
}
\end{figure}

The mechanism with which some points are first order, while some others are second order, is known in $d=2$ to be related to
a collision of fixed points in the RG diagram thanks to the explicit results using CFT methods \cite{Gorbenko:2018ncu,Gorbenko:2018dtm}.
The collision involves two fixed points, the standard critical one and a multicritical one,
that merge and annihilate each other into a complex pair as depicted in Fig.~\ref{fig:merging}.
We can realistically conjecture that this mechanism applies to the full diagram: in the region below the separatrix there
are two fixed points, critical and multicritical, while above there is no real fixed point. In other words,
the collision of the fixed points causes an \emph{effective} upper critical dimension $d_{\rm cl}(q)$.
Notice that the two fixed points might not be easily seen in the same RG diagram if perturbative methods are used
and the couplings controlling the respective perturbative series are of different canonical dimension, in which
case the annihilation of the fixed points would be visible in the respective RG diagrams as a fixed point going to infinity,
that is, the strong coupling regime.
In fact, this is precisely what happened to the critical point in $d=6-\epsilon$ when $q\to \frac{10}{3}$ [the coupling diverges as $(q-\frac{10}{3})^{-\nicefrac{1}{2}}$].

A candidate fixed point for the multicritical model that we want to push forward is our analytically continued ${\rm FP}_2$,
because it fits several of the expected properties that such multicritical point should have.
First, it is the most IR relevant fixed point for each value of $q$ that it exists.
Second, the only natural values of $q$ for which we always have a fixed point are $q=0$ and $q=1$.
Third, we know from the expansion in $d=\nicefrac{10}{3}-\epsilon$ that its maximum value is $q=q_2=1.8940$,
therefore it ceases to exist at the point $(\nicefrac{10}{3},1.8940)$, which is close enough to the $d=3$ line
to argue that $(3,3)$ lies in the first order region.
Finally, it ceases to exist by going to infinity, so in the strong coupling regime, in which
only through nonperturbative methods one would be able to observe the actual merging.

If our hypothesis is correct, the multicritical theory would provide a new analytically determined point
through which the separatrix should pass, giving a new valuable information on the Potts $(d,q)$ diagram.
Of course our hypothesis should be checked somehow, and the most natural way to do it would be
to use a nonperturbative RG method such as the functional renormalization group.
We therefore hope that the proof of this hypothesis is addressed by somebody in a future publication.

\paragraph*{Acknowledgements.}
For performing the computations of this paper we relied in part on the \emph{Mathematica}
packages \cite{xact-package,xperm-package} and \cite{Nutma:2013zea}.
OZ is grateful to R.~Ben Al\`i Zinati
for comments, discussions and the help provided with TikZ diagrams.

\appendix

\section{A summary of RG results and critical properties}\label{sect:relevant-formulas}

The general structure of $S_q$ invariant operators has been discussed in \cite{Vasseur:2013baa,Couvreur:2017inl} on general CFT grounds.
Following that result, the relevant operators that are multiplicatively renormalized
have been discussed in \cite{Safari:2020eut} for the critical model in $d=6-\epsilon$.
The same operators are scaling ones in this paper, so we follow the notation of \cite{Safari:2020eut},
from which we borrow the notation that $S$ stands for singlet, $V$ for vector, $T$ for $2$-tensor or simply tensor, and
$Z$ for $3$-tensor. We also use an apex to indicate how many copies of the field are needed to construct them.
The exact form of the scaling operators constructed with two fields ($S^{(2)}$, $V^{(2)}$ and $T^{(2)}$),
and the scaling operators constructed with three fields ($S^{(3)}$, $T^{(3)}$ and $Z^{(3)}$) can be found in \cite{Safari:2020eut}.
These six operators are scaling operators for each value of $q$, or, in other words, the action of dilatations is already diagonalized
and they do not mix through renormalization.

\begin{widetext}

The list of quadratic operators is a complete basis for symmetric quadratic deformations $\phi_i\phi_j$, because the only missing
irrep of $S_q$ is the antisymmetric one, that is obviously not realized without derivatives.
At the cubic level the above list is instead incomplete, because there are also two vectors, $V^{(2)}$ and $V^{(2)\prime}$,
that arise from diagonalizing
\begin{equation}\label{eq:mixing-cubic}
\begin{split}
 \sum_i \int {\rm d}^d x \Bigl\{ \frac{1}{2} \phi_i \phi^2 {\cal J}^i_{3,1} - \sum_{jkl} q^{(4)}_{ijkl} \phi_{j}\phi_{k}\phi_{l} {\cal J}^i_{3,2}
 \Bigr\}\,,
\end{split}
\end{equation}
in which we introduced two vector sources $ {\cal J}^i_{3,1}$ and $ {\cal J}^i_{3,2}$ that are mixed by renormalization.
We do not attempt the full generalization of the scaling analysis to the quartic level in the fields. However,
we give the scalar subsector, that is responsible for two scaling operators $S^{(4)}$ and $S^{(4)\prime}$
\begin{equation}\label{eq:mixing-quartic}
\begin{split}
 \delta S[\phi] &= 
 \int {\rm d}^d x \Bigl\{\lambda_{4,1} (\phi^2)^2 + \lambda_{4,2} \sum_{ijkl} q^{(4)}_{ijkl}\phi_{i}\phi_{j}\phi_{k}\phi_{l}
 \Bigr\}\,,
\end{split}
\end{equation}
for which it is sufficient to introduce two scalar sources $\lambda_{4,1}$ and $\lambda_{4,2}$.

A set of relevant operators ${\cal O}(\phi)$ that do not involve derivatives of the fields can be renormalized by
performing the replacement $V(\phi) \to V(\phi)+{\cal J}_{\cal O}\cdot {\cal O}(\phi)$ and renormalizing the sources ${\cal J}_{\cal O}$.
If we treat them as composite operators, the replacement requires the linearization of the RG equation \eqref{eq:flow-potential},
so the sources are renormalized mutiplicatively. If one is willing to go beyond the linear level, however, the additional operators can be
treated as full deformations of the potential (as long as they are relevant operators) and the sources acquire
fully-fledged beta functions \cite{Codello:2017hhh}. These are useful because the coefficients of these beta functions allow for the
determination of the coefficients of the operator product expansion (OPE). 
See also the leading order CFT results in~\cite{Codello:2018nbe} where quadratic operators were studied and some OPE coefficients extracted.

In this appendix, we give the renormalization of composite relevant operators in terms of gamma functions,
as a useful bridge from RG and CFT methods.
Given the gamma function $\gamma_{{\cal O}}$ of a scaling operator ${\cal O}$, that contains $n$ copies of the field $\phi$ and no derivatives,
the critical exponent $\theta_{{\cal O}}$ and the scaling dimension $\Delta_{{\cal O}} $ are
\begin{equation}
\begin{split}
  \theta_{{\cal O}}  = d-n\left(\frac{d-2+\eta}{2}\right) + \gamma_{{\cal O}}\,,
  \qquad\qquad
  \Delta_{{\cal O}} = n\left(\frac{d-2+\eta}{2}\right) - \gamma_{{\cal O}}\,.
\end{split}
\end{equation}
In the latter formula we assume that ${\cal O}$ is a primary operator,
because if it is a descendant the formula is adjusted to ensure that scaling dimensions
are consistent with the equations of motion $\Box \phi \sim V'(\phi)$, therefore $2+\Delta_\phi=\Delta_{V'}$.
From the RG point of view, the scaling relation $\theta_\phi+\theta_{V'}=d$ is always true
and can be proven in general.
Obviously, if the operators do not diagonalize the action of dilatations, like in \eqref{eq:mixing-cubic}
and \eqref{eq:mixing-quartic}, then we have a $\gamma$ matrix that requires further diagonalization.

The most important critical exponents are the anomalous dimension of the field $\eta/2$,
and the exponent of the scaling of the correlation length $\nu$.
The first important relation is the one that determines $\nu$,
that is identified with the inverse of the critical exponent of $S^{(2)} \sim \phi^2$ (the energy)
\begin{equation}
\begin{split}
 \nu^{-1}=\theta_{S^{(2)}}=2+\gamma_{S^{(2)}}-\eta\,.
\end{split}
\end{equation}
Other interesting exponents can be found for the rest of the $n=2$ sector, because they are identified with the fractal dimensions
\begin{equation}
\begin{split}
  d_{f} = \theta_{V^{(2)}}=2+\gamma_{V^{(2)}}-\eta 
  \,, \qquad\qquad
  d_{r} = \theta_{T^{(2)}}=2+\gamma_{T^{(2)}}-\eta
  \,.
\end{split}
\end{equation}
The dimension $d_f$ is the fractal dimension of propagator lines for the field theory \cite{Kompaniets:2019zes},
and therefore can be interpreted as the fractal dimension
of the clusters, since they live in the reciprocal space. The dimension $d_r$ has a similar meaning, but it is related to the property of the cluster
when seen as a resistivity network \cite{harris-fisch,dasgupta-et-al}.
Nonsinglet deformation generally lead to a breaking of the symmetry from $S_q$ to a subgroup, which often ends up to $\mathbb{Z}_2$
in a type of crossover phenomenon \cite{Bonati:2010ce}.
Close to the critical temperature, they also have a critical behavior
(similarly to the thermodynamical exponent $\delta$ in the case of the Ising model).
The critical exponents governing this behavior are called crossover exponents and are defined as
\begin{equation}
\begin{split}
  \overline{\Phi} = \frac{\theta_{V^{(2)}}}{\theta_{S^{(2)}}} = \frac{2+\gamma_{V^{(2)}}-\eta}{2+\gamma_{S^{(2)}}-\eta}
  \,, \qquad\qquad
  \Phi = \frac{\theta_{T^{(2)}}}{\theta_{S^{(2)}}}= \frac{2+\gamma_{T^{(2)}}-\eta}{2+\gamma_{S^{(2)}}-\eta}\,.
\end{split}
\end{equation}
Crossover exponents for the standard critical universality class and their scaling relations have been
discussed in \cite{wallace-young,Stephen:1977mw,Barbosa:1986kv,Theumann:1985qc}.
A summary of all critical exponents for the limits $q\to 0 $ and $q\to 1$ appears in Tables~\ref{tab:percolation} and \ref{tab:forest}, respectively.
\begin{table*}[htb]
\begin{tabular}{|l|l|l|l|l|l|l|l|}
\hline
 $q\to1$ 		& $\eta$ 					& $\nu^{-1}$ 				& $d_f$  					& $d_r$  					& $\Phi$  	& $\overline{\Phi}$ 			& $\alpha_E$  				\\ \hline
 ${\rm FP}_1$ 	& $-0.000219126\epsilon$  	& $2 + 0.00460164\epsilon$ 	& $2 + 0.0121628\epsilon$ 	& $2 + 0.00460164\epsilon$ 	& $1$ 	& $1 + 0.00378056\epsilon$	& $-0.0069779 \epsilon$ 	\\ \hline
 ${\rm FP}_2$ 	& $-0.00431785\epsilon$  	& $2 + 0.0906748\epsilon$ 	& $2 + 0.11886\epsilon$ 	& $2 + 0.0906748\epsilon$ 	& $1$ 	& $1 + 0.0140927\epsilon$	& $-0.0222583\epsilon$ 	\\ \hline
\end{tabular}
\caption{Summary of all critical exponents and properties for the two fixed points of the limit $q\to 1$ of Sect.~\ref{sect:percolation}.}
\label{tab:percolation}
\end{table*}
\begin{table*}[htb]
\begin{tabular}{|l|l|l|l|l|l|l|}
\hline
 $q\to0$ 		& $\eta$ 					& $\nu^{-1}$ 				& $d_f$  					& $d_r$  					& $\Phi$  					& $\overline{\Phi}$ 			\\ \hline
 ${\rm FP}_1$ 	& $-0.000508906\epsilon$  	& $2 + 0.010687\epsilon$ 	& $2 + 0.0183206\epsilon$ 	& $2 + 0.00335953\epsilon$ 	& $1 - 0.00366375\epsilon$ 	& $1 + 0.00381679\epsilon$	\\ \hline
 ${\rm FP}_2$ 	& $-0.000508906\epsilon$  	& $2 + 0.010687\epsilon$ 	& $2 + 0.0183206\epsilon$ 	& $2 + 0.0816582\epsilon$ 	& $1 + 0.0354856\epsilon$ 	& $1 + 0.00381679\epsilon$	\\ \hline
 ${\rm FP}_3$ 	& $-0.000508906\epsilon$  	& $2 + 0.010687\epsilon$ 	& $2 + 0.0183206\epsilon$ 	& $2 + 0.0189587\epsilon$ 	& $1 + 0.00413585\epsilon$ 	& $1 + 0.00381679\epsilon$	\\ \hline
 ${\rm FP}_4$ 	& $0$  					& $2$ 		 			& $2$ 					& $2 + 0.0434783$ 		& $1 + 0.0217391\epsilon$ 	& $1$					\\ \hline
\end{tabular}
\caption{Summary of all critical exponents and properties for the four fixed points of the limit $q\to 0$ of Sect.~\ref{sect:forest}.}
\label{tab:forest}
\end{table*}

Now we collect all the gamma functions that have been determined for this work.
For the quadratic operators
\begin{equation}\label{eq:gamma-quadratic}
\begin{split}
 \gamma_{S^{(2)}} &=
  -\frac{1}{15} {u}^2 (q-2) (q+5)
  -\frac{4}{3} {u} {v} (q-2) (q-1)
  -\frac{2}{3} {v}^2 (q-2) \left(q^2-2 q+2\right) \,,
  \\
 \gamma_{V^{(2)}} &=
  \frac{1}{150} {u}^2 \left(-4 q^2-57 q+175\right)
  -\frac{2}{15} {u} {v} (2 q-5) (5 q-7)
  -\frac{2}{3} {v}^2 \left(q^3-5 q^2+9 q-7\right) \,,
  \\
 \gamma_{T^{(2)}} &=
  \frac{1}{150} {u}^2 \left(-q^2-12 q+73\right)
  -\frac{2}{15} {u} {v} (q-5) (q-1)
  +\frac{2 {v}^2}{3} \,.
\end{split}
\end{equation}
For the cubic scaling operators
\begin{equation}
\begin{split}
 \gamma_{S^{(3)}} &=
  \frac{1}{100} {u}^2 \left(19 q^2+399 q-1000\right)
   +\frac{1}{5} {u} {v} \left(61 q^2-219 q+200\right)
   +{v}^2 \left(7 q^3-35 q^2+63 q-40\right) \,,
  \\
 \gamma_{T^{(3)}} &=
  \frac{1}{100} {u}^2 \left(-q^2+157 q-738\right)
  +\frac{1}{5} {u} {v}\left(q^2-17 q+58\right)
  +\frac{1}{3} {v}^2 \left(-2 q^3+14 q^2-21 q-6\right) \,,
  \\
 \gamma_{Z^{(3)}} &=
  \frac{1}{50} {u}^2 \left(-q^2+6 q-71\right)
  -\frac{2}{5} {u} {v} (q-5) (q-1)
  +2 {v}^2 \,.
\end{split}
\end{equation}
The mixing matrix of \eqref{eq:mixing-cubic} that leads to $V^{(3)}$ and $V^{(3)\prime}$ is
\begin{equation}
\begin{split}
 \gamma_{V^{(3)}} &=
  \left(
\begin{array}{cc}
   \gamma^{V^{(3)}}_{11}
   & \gamma^{V^{(3)}}_{12} \\
   \gamma^{V^{(3)}}_{21}
   &  \gamma^{V^{(3)}}_{22} 
\end{array}
\right) \,,
 \\
 \gamma^{V^{(3)}}_{11}
 &=
 \frac{1}{100} {u}^2 \left(q^2+543 q-1126\right)
 +\frac{1}{10} {u} {v} \left(131 q^2-483 q+544\right)
 +{v}^2 \left(7 q^3-35 q^2+72 q-58\right) \,,
 \\
 \gamma^{V^{(3)}}_{12}
 &=
 -\frac{3}{25} {u}^2 \left(19 q^2-141 q+212\right)
 -\frac{3}{5} {u} {v} \left(3 q^3-27 q^2+63 q-76\right)
 +2 {v}^2 \left(2 q^2+3 q-12\right) \,,
 \\
 \gamma^{V^{(3)}}_{21}
 &=
 \frac{1}{25} {u}^2 (-4 q-41)
 +\frac{1}{20} {u} {v} \left(-3 q^2-111q+188\right)
 +\frac{1}{6} {v}^2 \left(-23 q^2+69 q-60\right) \,,
 \\
 \gamma^{V^{(3)}}_{22}
 &=
 \frac{1}{100} {u}^2 \left(7 q^2+73 q-566\right)
 +\frac{1}{10} {u} {v} (q-1) (5 q-56)
 -\frac{2}{3} {v}^2 \left(q^3-4 q^2+6 q+3\right) \,.
\end{split}
\end{equation}
The mixing matrix of \eqref{eq:mixing-quartic} that leads to $S^{(4)}$ and $S^{(4)\prime}$ is
\begin{equation}
\begin{split}
 \gamma_{S^{(4)}} &=
  \left(
\begin{array}{cc}
   \gamma^{S^{(4)}}_{11}
   & \gamma^{S^{(4)}}_{12} \\
   \gamma^{S^{(4)}}_{21}
   &  \gamma^{S^{(4)}}_{22} 
\end{array}
\right) \,,
 \\
 \gamma^{S^{(4)}}_{11}
 &=
 \frac{3}{100} {u}^2 \left(13 q^2+150 q-1051\right)
 +\frac{1}{5} {u} {v} (q-1) (25 q-181)
 +\frac{1}{3} {v}^2 \left(-4 q^3+7 q^2-6 q-75\right) \,,
 \\
 \gamma^{S^{(4)}}_{12}
 &=
 \frac{3}{100} {u}^2 \left(q^2+56 q+583\right)
 +\frac{1}{15} {u} {v} \left(23 q^2+900 q-1499\right)
 +\frac{1}{3} {v}^2 \left(119 q^2-348 q+311\right) \,,
 \\
 \gamma^{S^{(4)}}_{21}
 &=
 \frac{3}{100} {u}^2 \left(3 q^3+166 q^2-1358 q+2149\right)
 +\frac{1}{5} {u} {v} \left(23 q^3-206 q^2+498 q-603\right)
 +{v}^2 \left(-4 q^2-42 q+87\right) \,,
 \\
 \gamma^{S^{(4)}}_{22}
 &=
 \frac{1}{100} {u}^2 \left(29 q^2+2886 q-6251\right)
 +\frac{1}{5} {u} {v} \left(363 q^2-1342q+1479\right)
 +\frac{1}{3} {v}^2 \left(115 q^3-575 q^2+1170 q-921\right) \,.
\end{split}
\end{equation}

We conclude by briefly explaining the logarithmic structure that appears in the limit $q\to1$, summarizing \cite{vjs},
to which we remind for many more details.
The scaling dimensions of the operators $S^{(2)}$ and $T^{(2)}$ are degenerate in the limit $q\to1$,
so they form a logarithmic pair for the case of percolations.
Physically, they govern the leading nontrivial behavior of the energy, $E\sim S^{(2)}$, and of the $2$-cluster, $\tilde{E}\sim T^{(2)}$, operators.
The general $q$-dependent form of the correlators for $q\sim 1$ is
\begin{equation}
 \begin{split}
 &\langle E(x) E(0) \rangle = (q-1) \frac{ A }{\left|x\right|^{2\Delta_E}}\,, \\
 &\langle \tilde{E}_{ij}(x) \tilde{E}_{kl}(0) \rangle =  \frac{2}{q^2}
  \Bigl(\delta_{ik}\delta_{jl}+\delta_{ik}\delta_{jl} -\frac{\delta_{ik}+\delta_{il}+\delta_{jk}+\delta_{jl}}{q-2}
  +\frac{2}{(q-1)(q-2)}\Bigr)
 \frac{A}{\left|x\right|^{2\Delta_{\tilde{E}}}}\,,
 \end{split}
\end{equation}
given that the normalizations are constrained by the requirement of $S_q$ symmetry \cite{vjs}.
The limit $q\to 1$ is singular because the two operators
fall into the same Jordan cell and the dilatations are not diagonalized anymore, $\Delta_E=\Delta_{\tilde{E}} \equiv\Delta$ for $q\to 1$.
In the limit, we keep the energy operator, that scales as $\Delta$, but also define a new operator
\begin{equation}
 \begin{split}
 \hat{E}_{ij}(x) \equiv \tilde{E}_{ij}(x) +\frac{2}{q(q-1)} E(x)\,.
 \end{split}
\end{equation}
The requirement that the correlator of $E(x)$ is regular in the limit $q\to 1$
fixes $A=A(q)$, and the energy behaves as a normal scaling operator.
Using $\hat{E}_{ij}(x)$, one finds a regular $q\to 1$ limit for the correlator
\begin{equation}
 \begin{split}
 &\langle \hat{E}_{ij}(x) \hat{E}_{kl}(0) \rangle = 2
  \Bigl(\delta_{ik}\delta_{jl}+\delta_{ik}\delta_{jl} 
 +\delta_{ik}+\delta_{il}+\delta_{jk}+\delta_{jl}
  +4 \alpha_E \log \left|x\right| \Bigr)
 \frac{A}{\left|x\right|^{2\Delta}}\,,
 \end{split}
\end{equation}
that, besides the leading scaling as $\Delta$, has a logarithmic term, differently from the energy. Interestingly, the coefficient $\alpha_E$ of the logarithm is universal
\begin{equation}\label{eq:alpha-universal}
 \begin{split}
\alpha_E \equiv \lim_{q\to 1} \frac{\Delta_{\tilde{E}}-\Delta_E}{q-1}
 =-\frac{6 {{u}}^2}{25 }+\frac{4 {{u}} {{v}}}{5 }-\frac{2 {{v}}^2}{3 }
 \end{split}
\end{equation}
because it is determined as the limit of the difference of scaling dimensions that are also universal.
In the last step, we used \eqref{eq:gamma-quadratic} and explicitely performed the limit to make a connection with the multicritical model of the paper \eqref{eq:action};
the couplings are understood to be at one of the fixed points of Sect.~\ref{sect:percolation}.
The coefficient $\alpha_E$ can be measured on critical percolations through an opportune observable
\cite{vjs,Vasseur:2013baa,Couvreur:2017inl,Gori:2017cyq}.

\end{widetext}

\bibliography{bibliography}

\end{document}